\documentclass{ws-procs9x6}
\usepackage{graphicx}

\newcommand{\intvecx}{\int d^3 x\,}

\newcommand{\vecA}{{\bf A}}

\newcommand{\veck}{{\bf k}}

\newcommand{\vecx}{{\bf x}}       

\newcommand{\al}{\alpha}
\newcommand{\bt}{\beta}
\newcommand{\gm}{\gamma}
\newcommand{\dl}{\delta}
\newcommand{\ep}{\epsilon}

\newcommand{\kp}{\kappa}
\newcommand{\lm}{\lambda}
\newcommand{\rh}{\rho}
\newcommand{\sg}{\sigma}

\newcommand{\ph}{\phi}
\newcommand{\vr}{\varphi}

\newcommand{\om}{\omega}

\newcommand{\half}{\frac{1}{2}}
\newcommand{\quart}{\frac{1}{4}}
\newcommand{\third}{\frac{1}{3}}
\newcommand{\Tr}{\mbox{Tr}}
\newcommand{\tr}{\mbox{tr}}

\newcommand{\dm}{\partial_{m}}

\newcommand{\dt}{\partial_{t}}

\newcommand{\eela}[1]{\label{#1}\end{equation}}
\newcommand{\eeala}[1]{\label{#1}\end{eqnarray}}
\newcommand{\be}{\begin{equation}}
\newcommand{\ee}{\end{equation}}
\newcommand{\bea}{\begin{eqnarray}}
\newcommand{\eea}{\end{eqnarray}}
\newcommand{\vrd}{\vr^{\dagger}}
\newcommand{\ncs}{N_{\rm CS}}
\newcommand{\nw}{N_{\rm w}}

\newcommand{\lmsp}{\lm_{\sg\ph}}
\newcommand{\mueff}{\mu_{\rm eff}}
\newcommand{\beaa}{\begin{eqnarray*}}
\newcommand{\eeaa}{\end{eqnarray*}}
\newcommand{\bc}{\begin{center}}
\newcommand{\ec}{\end{center}}
\newcommand{\dlcp}{\delta_{\rm CP}}
\newcommand{\gmeff}{\Gamma_{\rm eff}}
\begin{document}

\title{Cold Electroweak Baryogenesis
}

\author{Jan Smit}

\address{Institute for Theoretical Physics, University of Amsterdam\\
Valckenierstraat 65, 1018 XE Amsterdam, the Netherlands}

\maketitle

\abstracts{We present arguments that the CKM CP-violation in the standard model 
may be sufficient
for the generation of the baryon asymmetry, if the electroweak transition
in the early universe was of the cold, tachyonic, type
after electroweak-scale inflation. A model implementing this is
described which complies with the CMB constraints and which is falsifiable with
the LHC. 
Numerical simulations of the tachyonic transition 
with an effective CP-bias indicate that the observed
baryon asymmetry can be generated this way.
}

\section{Introduction}
There are three families in the minimal Standard Model (SM), why?
Such why-questions lie outside the scope of the model, but the thought
`to allow for CP-violation and baryogenesis' comes up easily. What is
currently known experimentally about CP violation\cite{pdg} can be
attributed to the CKM matrix being complex, which is only
possible for three or more families.
This provides a strong
motivation for finding a scenario for baryogenesis that makes use of
this fact.  SM electroweak baryogenesis\cite{Rubakov:1996vz},
is believed by many to be impossible, because of ($i$) the
weakness of CKM CP-violation, and ($ii$) the smoothness of the 
finite-temperature electroweak transition\cite{Kajantie:1995kf}.

Of course, the SM assumes neutrinos to be massless, and a better framework
is its extension that includes right-handed neutrino fields
with renormalizable Yukawa couplings and a Majorana mass matrix,
which may be called the Extended Standard Model (ESM). Leptogenesis is
an attractive possibility within this framework\cite{Plu}. However,
it makes use of physics at the scale of $10^{10}$ GeV and it seems worth
putting effort into scenarios closer to what we know on the scale $\lesssim 200$
GeV. So we must face issues ($i$) and ($ii$).

In the following a scenario\cite{Garcia-Bellido:1999sv,Krauss:1999ng,Copeland:2001qw,Tranberg:2003gi,vanTent:2004rc}
will be reviewed based on a cold (`tachyonic')
electroweak transition caused by the coupling of a gauge-singlet inflaton to 
the ESM.
First, we shall
argue that CP violation is much stronger at {\em zero} temperature
than often stated.

\section{Magnitude of CKM CP-violation}
\label{strength}
The usual estimate for the magnitude of CP violation is\cite{Jarlskog:1985ht,Rubakov:1996vz}
\be
\dl_{\rm CP} = J\,
(m_{u}^{2}-m_{c}^{2}) (m_{c}^{2}-m_{t}^{2}) (m_{t}^{2}-m_{u}^{2})
(m_{d}^{2}-m_{s}^{2}) (m_{s}^{2}-m_{b}^{2}) (m_{b}^{2}-m_{d}^{2})
/T^{12},
\label{dlcpest1}
\ee
where $m_u$, \ldots,$m_b$ are the quark masses and\cite{pdg}
\be
J = |{\rm Im} (V_{fg}V_{hi}V_{fi}^* V_{hg}^*)| = (2.88 \pm 0.33)
\times 10^{-5}
\label{J}
\ee
is the simplest rephasing-invariant combination of the CKM matrix $V$. 
For $T=100$ GeV, of order of
the finite-temperature electroweak phase transition, (\ref{dlcpest1}) gives
the discouraging value $\dlcp \approx 10^{-19}$.
At zero temperature one would replace $m_j/T$ by the Yukawa
coupling $\lm_j = \sqrt{2}\, m_j/v$, with $v= 246$ GeV the vev of
the Higgs field, giving
\be
\dl_{\rm CP} = J\,
(\lm_{u}^{2}-\lm_{c}^{2}) (\lm_{c}^{2}-\lm_{t}^{2})
(\lm_{t}^{2}-\lm_{u}^{2}) (\lm_{d}^{2}-\lm_{s}^{2})
(\lm_{s}^{2}-\lm_{b}^{2}) (\lm_{b}^{2}-\lm_{d}^{2})
\approx 10^{-22},
\label{dlcpest2}
\ee
even smaller than (\ref{dlcpest1}).
However, even with the Higgs field settled in its vev, the
measured CP violating effects in accelerator experiments are at a
much higher level than $10^{-23}$, e.g.\ the magnitude of the
decay asymmetry $\ep'$ in the $K^0-\bar K^0$ system is about\cite{pdg}
$4\times 10^{-6}$.
This suggests that the above order of magnitude estimates of
$\dl_{\rm CP}$ are misleading, at least at zero temperature.

To make the discussion more concrete, consider the effective action obtained
by integrating our the fermions, $\gmeff = -\Tr(\ln D)$,
with $D$ the Dirac operator. In studies and discussions of
electroweak baryogenesis,
CP violation has been taken into account by approximating the
CP asymmetry in $\gmeff$ by a leading dimension-six term\cite{Shaposhnikov:1987tw}
\be
-\mathcal{L}_{\rm CP} = 
\frac{3\delta_{\rm CP}}{16\pi^2 M^2}\,
\vr^{\dagger}\vr\; \tr(F^{\mu\nu}\tilde{F}_{\mu\nu}),
\label{CPterm}
\ee
where $F_{\mu\nu}$ is the SU(2) field strength tensor,
$\tilde F_{\mu\nu}$ its dual, and
$M$ is a mass depending on the scale of the problem. 
It could be the mass scale of an extension of the ESM. 
{\em Within} the ESM, at finite temperature,
one could take $M=T$, but what to use when $T=0$?
The fermion masses are $\propto \langle \vr\rangle$,
but $\vrd\vr/\langle\vrd\vr\rangle$ does not seem to make sense.
The ordering of the terms in $\gmeff$ according to increasing
dimension is questionable in a zero-temperature
transition in which $\langle\vr^{\dagger}\vr\rangle$ increases from zero
to some finite value.

Luckily, there exists a very detailed calculation of the zero-temperature
effective action in a general chiral gauge model,
based on a gauge-covariant derivative-expansion that is
completely non-perturbative in the Higgs field, by L.L.\ Salcedo\cite{Salcedo:2000hp,Salcedo:2000hx}.
We have applied his results to the ESM (without Majorana mass terms), and
found that (\ref{CPterm}) is incorrect for this case:
the rephasing invariant $J$ does not appear as a coefficient of
$\tr(F^{\mu\nu}\tilde F_{\mu\nu})$ times a function of the un-differentiated
Higgs field\cite{Smit:2004kh}.
The first CKM CP-violating contribution
lies unfortunately still beyond the scope of Salcedo's
results,
but a typical term is expected to have the form, in {\em unitary gauge}
$A_{\mu} \to W_{\mu}$, 
$\vr = (0,h)^T$,
\[
\ep^{\kp\lm\mu\nu} W_{\kp\lm}^a W_{\mu\nu}^b W^c_{\rh} W^{d\rh}\, n'_{abcd}(\lm h),
\]
where $a,b,c,d$ are $SU(2)$ indices and $n'_{abcd}(\lm h)$ is a non-trivial
function of 
the Yukawa couplings times the Higgs field,
$d_j \equiv \lm_j\, h$.
For e.g.\ $a=b=c=d=1$ the above expression violates CP and is 
expected to contain the invariant $J$. The crucial point is now that
$n'_{abcd}$ is expected to be a {\em homogeneous} function of the $d_j$. 
This is the case for the explicitly calculated 
coefficient functions at fourth order of the 
gauge-covariant derivative-expansion (involving four Lorentz indices),
which are homogeneous of degree zero\cite{Salcedo:2000hp,Salcedo:2000hx},
which is why (\ref{CPterm}) cannot occur.
For the higher order term we expect that,
when we rescale the Yukawa couplings $\lm_j\to s \lm_j$, 
the coefficient functions $n(\lm h)$
scale by some negative power of $s$ (perhaps up to logarithms). 
This strongly suggests
that we should not include the product of Yukawa couplings (\ref{dlcpest2})
in estimating $\dlcp$, leaving only $J$, which is a factor of about $10^{17}$
larger.
For example,
\[
\frac{(\lm_{u}^{2}-\lm_{c}^{2}) (\lm_{c}^{2}-\lm_{t}^{2})
(\lm_{t}^{2}-\lm_{u}^{2}) (\lm_{d}^{2}-\lm_{s}^{2})
(\lm_{s}^{2}-\lm_{b}^{2}) (\lm_{b}^{2}-\lm_{d}^{2})}
{(\lm_{u}^{2}+\lm_{c}^{2}) (\lm_{c}^{2}+\lm_{t}^{2})
(\lm_{t}^{2}+\lm_{u}^{2}) (\lm_{d}^{2}+\lm_{s}^{2})
(\lm_{s}^{2}+\lm_{b}^{2}) (\lm_{b}^{2}+\lm_{d}^{2})}
\simeq 0.99.
\]
The argument applies only at zero temperature,
since at finite $T$ a new scale appears, e.g.\ the thermal QCD
quark mass $m_{\rm th} = g_{\rm s} T/\sqrt{6}$.

\section{A falsifiable model of electroweak-scale inflation}
\label{falsif}
Here we take seriously a minimal phenomenological
extension of the ESM with an additional 
gauge-singlet inflaton that couples only to the Higgs field\cite{vanTent:2004rc}.
Its inflaton-Higgs potential is constructed after \cite{Copeland:2001qw}, 
\be
V(\sigma,\ph) = V_0 - \frac{1}{p} \al_p \sigma^p + \frac{1}{q}
\al_q \sigma^q
- \frac{1}{2} \lmsp \sigma^2 \ph^2 + \frac{1}{2} \mu_\ph^2 \ph^2 +
\frac{1}{4} \lm_\ph (\ph^2)^2,
\label{core}
\ee
where $\sg$ is the inflaton field and $\ph^2 \equiv 2\vrd\vr$.
During inflation, $\sg$ slow-rolls away from the origin. After
inflation has ended it accelerates, and
the effective Higgs mass
\be
\mueff^2 = \mu_\ph^2 - \lmsp \sg^2
\ee
changes sign from positive to negative, and the tachyonic electroweak
transition starts. During the transition the baryon asymmetry is
to be generated through the electroweak anomaly and the
CKM CP-bias. To avoid sphaleron washout of the asymmetry the
reheating temperature
$T_{\rm rh}$ has to be small enough. Taking $V_0^{1/4}=100$ GeV,
the Hubble rate is tiny, $H\approx 10^{-14}$ GeV, 
and $T_{\rm rh} =[30 V_0/(\pi^2 g_*)]^{1/4} \simeq
51$ GeV ($g_* = 86.25$ is the effective number of SM degrees of
freedom below the $W$ mass), at which temperature the sphaleron
rate is negligible.

As a phenomenological model it has to comply with what is
currently known. Firstly, we have the CMB constraints\cite{Spergel:2003cb}: 
the large-scale normalization
$|\dl_\veck|^2 = (3.8 \pm 0.5)\times 10^{-5}$, and the scalar spectral-index
$n_S-1 = -0.03\pm 0.03$. The latter is given by $n_S-1 \simeq -
(2/N)(p-1)/(p-2)$, with $N$ the number of e-folds before the end
inflation, and $p>2$ for slow-roll to end naturally. For
electroweak-scale inflation
$N\simeq 24$, which is much lower than usually mentioned $\approx
60$, and the resulting
$n_S-1 \lesssim -0.083$ is too low. This is fixed\cite{vanTent:2004rc} 
as mentioned below.

Secondly, in the minimum of the potential we have (approximating
the cosmological constant by zero)
\bea
\frac{dV}{d\sg}(v_\sg,v_\ph)&=&0,\;\;\frac{dV}{d\ph}(v_\sg,v_\ph)=0,
\quad V(v_\sg,v_\ph)=0,
\quad v_\ph = 246\; \mbox{GeV},
\nonumber\\
\mu_\ph^2 -\lmsp v_\sg^2 &=& -\half\, m_{\ph}^2,
\;\;
m_\ph^2 = 2\lm_\ph v_\ph^2,
\label{lmspconstraint}
\eea
where $m_\ph$ would be the SM Higgs mass, were it not that $\lmsp$
leads to a considerable inflaton-Higgs mixing, to be addressed
shortly. For definiteness we continue with $V_0^{1/4} = 100$ GeV,
$\mu_\ph = 100$ GeV, $m_\ph= 200$ GeV, for which the Higgs self-coupling
$\lm_\ph=0.33$, and we expect a particle mass above the current lower bound
of 114 GeV\cite{pdg}.

Thirdly, the transition has to be reasonably rapid to
allow for successful baryogenesis, 
which leads\cite{Copeland:2001qw,vanTent:2004rc} to the constraint 
$\lmsp \gtrsim 0.01$.
Essentially this constraint now fixes $p=5$ (the choices for $V_0$,
$\mu_\ph$ and $m_\ph$ are not critical in this respect). 
The basic reason is that for $p<5$ the potential goes down rather slowly, giving
a large $\sg$-vev 
$v_\sg$ and consequently a too small $\lmsp$
(cf.\ (\ref{lmspconstraint})), whereas for $p>5$
the result is a ridiculously large $\lmsp$.
%
So we end up with non-renormalizable couplings, 
and to keep the powers of $\sg$ as low as possible we have chosen $q=6$.

Returning to the minimum of the potential, the inflaton-Higgs
mixing follows from the mass matrix
\[
\partial^2 V/\partial(\sg,\ph)_{v_\sg,v_\ph}
\Rightarrow \mbox{$m_1 = 385$ GeV, $m_2 = 125$ GeV, $\xi = 0.44$
},
\]
where $\xi$ is the mixing angle. The model thus predicts (only)
two scalar particles with couplings to the rest of the SM
determined by the mixing.

\begin{figure}[h]
\centerline{\epsfxsize=6cm\epsfbox{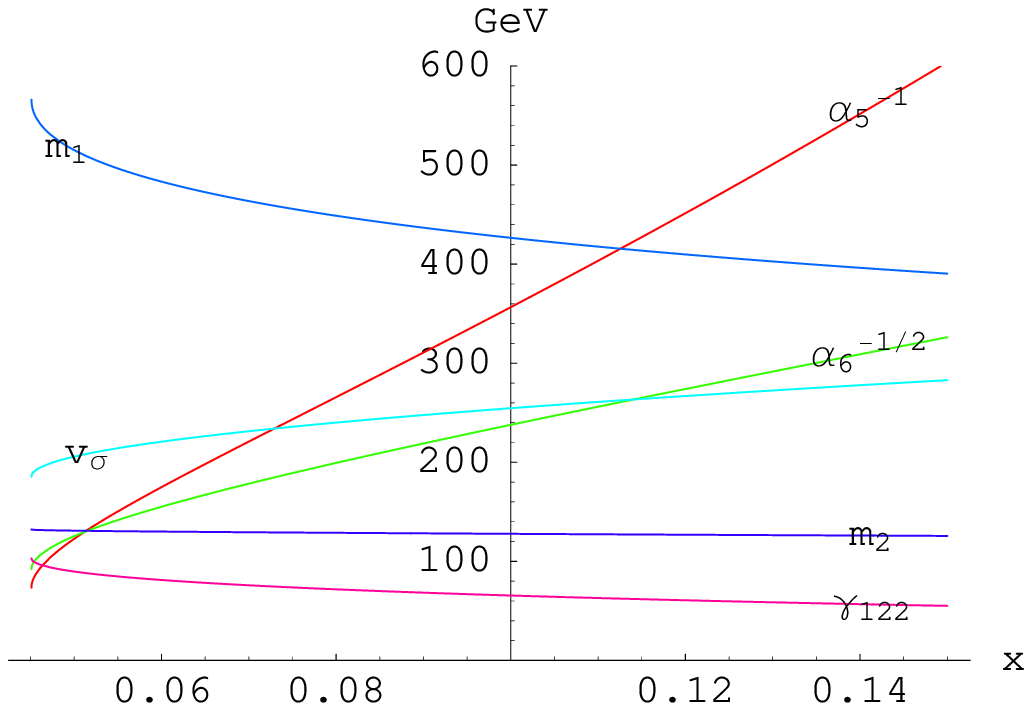}
\epsfxsize=6cm\epsfbox{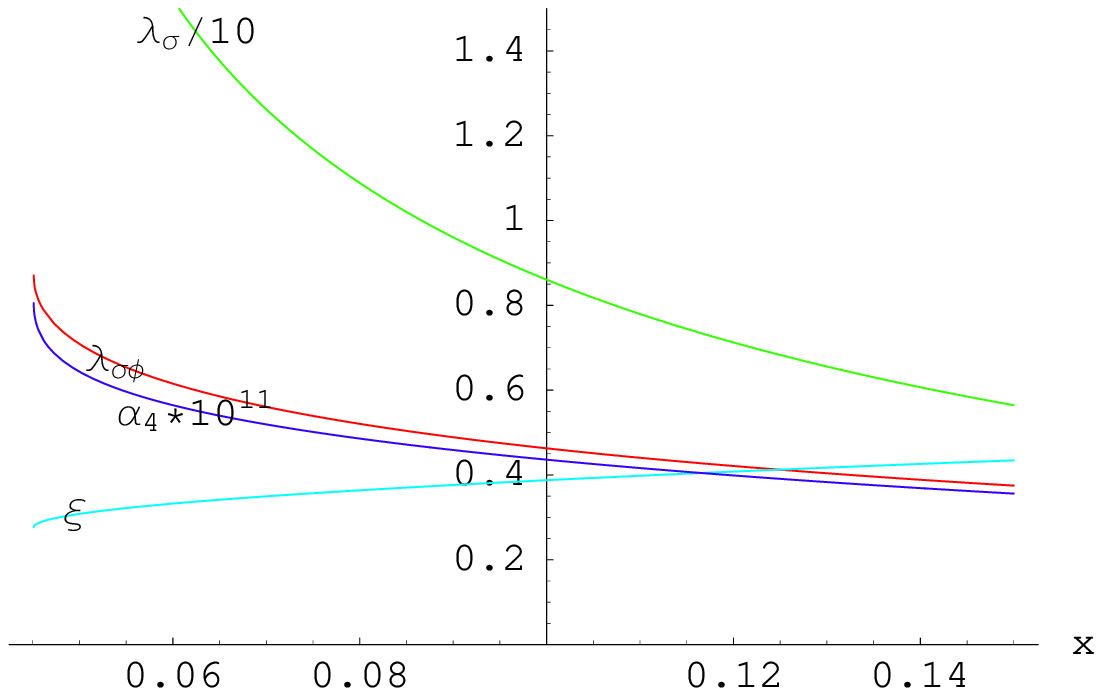}
%
}
\caption{Dimension-full (left) and dimension-less (right) parameters
as a function of 
$x\equiv \al_2 m_{\rm Planck}^2/V_0$, for the central WMAP value
$n_S-1=-0.03$.
\label{pars}}
\end{figure}

The CMB scalar spectral-index can be accommodated by adding terms
$-\half\,\al_2\sg^2 + \third\, \al_3 \sg^3 -\quart\, \al_4\sg^4$
to the potential that were artificially set to zero in (\ref{core}).
For example, still with $\al_3=0$ we found that we could
fit $n_S$ easily, see figure \ref{pars}.
Furthermore, in a one-loop study we verified that the potential
can be kept flat enough so as not to upset slow-roll inflation;
the non-renormalizable couplings lead to the conclusion that
the model breaks down at scales around a TeV or so.

The flatness of the potential is finely tuned, but we stress that
there is nothing wrong with this in a {\em phenomenological}
model. 
It is a minimal extension of the SM that incorporates
inflation as well as the tachyonic electroweak transition desired
for baryogenesis, 
with signatures due to inflaton-Higgs
mixing that allow it to be {\em falsified} by the LHC.

\section{
Tachyonic way-out-of-equilibrium transition}
\label{quench}

After inflation and before the transition, the universe is cold and
approximately in the ground state corresponding to the symmetric
phase of the SM. Then $\mueff^2$ changes sign and the tachyonic
transition develops. The initial part of such transitions with
given `speed' $\partial \mueff^2/\partial t$ has been studied
analytically as well as numerically\cite{Asaka:2001ez,Copeland:2002ku,Garcia-Bellido:2002aj,Garcia-Bellido:2003wd,Borsanyi:2003ib}.
We used the infinitely-fast quench,
\beaa
\mueff^2 &=& +\mu^2,\quad t<0
\\&=& -\mu^2, \quad t>0,
\eeaa
in our simulations of the $SU(2)$-Higgs 
model described by
\be
-\mathcal{L} = \frac{1}{2g^2}\,\tr(F_{\mu \nu}F^{\mu \nu})
 +(D_{\mu}\vr)^{\dagger}D^{\mu}\vr
 + \mueff^2 \vr^{\dagger}\vr+\lambda(\vr^{\dagger}\vr)^{2}
\label{su2hlag}
\ee
as an explorative approximation for studying baryogenesis\cite{Tranberg:2003gi}. 
In the gaussian approximation the Fourier modes of a real
component
$\ph_\veck$ of the Higgs field can be solved in the form
$\al_{\veck}\, e^{-i\omega_{k}^{-}t}
+ \bt_\veck\, e^{i\omega_{k}^{-}t}$, with
$\omega_{k}^{-}=\sqrt{-\mu^2 +k^{2}}$. The modes with
$k<\mu$ are unstable and grow exponentially $\propto e^{|\om_k^-|t}$.
After some time $\mu t \gg 1$, 
the equal-time correlators of $\ph_\veck$ and $\pi_\veck = \dt\ph_\veck$ 
are generically well approximated by the {\em classical distribution}\cite{Smit:2002yg}
\be
\exp\left[-\half\sum_{|\veck|<\mu }
\left(\frac{|\xi^+_\veck|^2}{n_k + 1/2 + \tilde n_k}
+ \frac{|\xi^-_\veck|^2}{n_k + 1/2 - \tilde n_k}\right)\right],
\label{dist}
\ee
where $\xi^{\pm}_{\veck} =
\frac{1}{\sqrt{2\om_k}}\left(\om_k\,\ph_{\veck} \pm
\pi_\veck\right)$,
and $n_k + 1/2 + \tilde n_k \gg 1$ and
$n_k + 1/2 - \tilde n_k \ll 1$. So the typical $\xi_k^+$ grow
large, while $\xi_k^- \to 0$; the distribution gets {\em
squeezed}. In the classical approximation\footnote{See \cite{ST}
for a comparison in the $O(4)$ model with the quantum case using 
$\Phi$-derivable approximation.}, realizations of the
distribution (\ref{dist}) are used as initial conditions for time
evolution using the classical equations of motion.
For a weakly coupled system such as the SU(2)-Higgs sector of the
SM we don't have to wait till the occupation numbers have grown
large, as the quantum and classical evolutions are identical in
the gaussian approximation. Our `just a half' initial conditions
start directly at $t=0$, for which the occupation numbers $n_k =\tilde n_k
=0$. In temporal gauge, $A_0=0$, the gauge field $\vecA$
is initialized to zero, and the initial $\dt\vecA$ is found from
Gauss' law\cite{Tranberg:2003gi}. 

\begin{figure}[h]
\bc
\includegraphics*[width=10cm,clip]{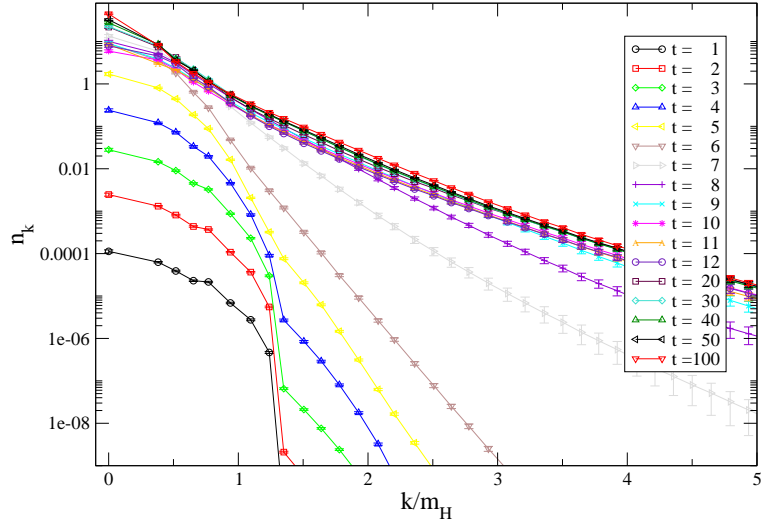}
\ec
\caption{Coulomb-gauge W-particle numbers vs.\ momentum, time in
units of $m_H^{-1}$. }
\label{figzeromod}
\end{figure}

In \cite{Skullerud:2003ki} 
we studied the particle-number distribution of the Higgs and W.  
For example,
Fig.\ \ref{figzeromod} shows how the W-particle $n_k$ in Coulomb gauge
change in time. The rapid rise of the low momentum modes end at
about $t m_H = 7$ and after
$t m_H =30$ the distribution does not change much anymore.
These $n_k$ differ initially very much from the corresponding ones
in the unitary gauge, but by time $t m_H = 100$ they are
practically the same. The particle numbers of
the longitudinal polarization are initially
similar to those of the Higgs but later appear to be slower in
adjusting to the same distribution. The distributions at $t m_H =
100$ can be fitted to Bose-Einstein distributions, separately in
the more important low momentum region and for the higher momentum
tails, giving effective temperatures and (at that time still
considerable) chemical potentials. The overall lesson learned from
these studies is that the transition is strongly out of equilibrium, and
over after a few tens of
$m_H^{-1}$. The finer details of thermalization will be different
when the other d.o.f.\ from the quarks, gluons, etc.\
have been taken into account.

\section{Baryogenesis}
To estimate the asymmetry generated by the transition under influence of
CP violation we simulated\cite{Tranberg:2003gi}
the $SU(2)$-Higgs model with the effective CP
violating term (\ref{CPterm}) added to the lagrangian (\ref{su2hlag}).
Although it was mentioned in section \ref{strength} that this term does not apply
to the ESM, it is still interesting to see how large its parameter
$\dlcp$ has to be in order to obtain the observed asymmetry. For the simulation
we relabelled
\be
\frac{3\delta_{\rm CP}}{16\pi^2 M^2} \equiv \kp,
\quad
k \equiv 16\pi^2 m_W^2\, \kp.
\label{kappa}
\ee
The baryon asymmetry is given by the anomaly equation
\[
B(t) = 3 \ncs(t) = 3 \frac{1}{16\pi^2}\int_0^t dt' \intvecx 
\left\langle \tr\left(F_{\mu\nu}(\vecx,t') \tilde F^{\mu\nu}(\vecx,t')\right) \right\rangle,
\]
where we assumed $B(0)=0$ (after inflation,
just before the transition) and initial Chern-Simon number
$\ncs(0)=0$ (choice of gauge). Another useful observable is the winding number
of the Higgs field,
\bea
\nw &=&\frac{1}{24\pi^2}\intvecx \ep_{klm}
\tr (\partial_k U U^{\dagger} \partial_l U U^{\dagger}
\dm U U^{\dagger}),
\nonumber\\
U &=&
\left(i\tau_2 \vr^*,\,\vr \right)/(\vrd\vr)^{1/2}
\in SU(2).
\nonumber
\eea
At relatively
low energy the gauge and Higgs fields are close to being pure-gauge and then 
$\ncs(A) \approx \nw(\vr)$.

\begin{figure}[h!]
\centerline{
\includegraphics*[width=8cm,clip]{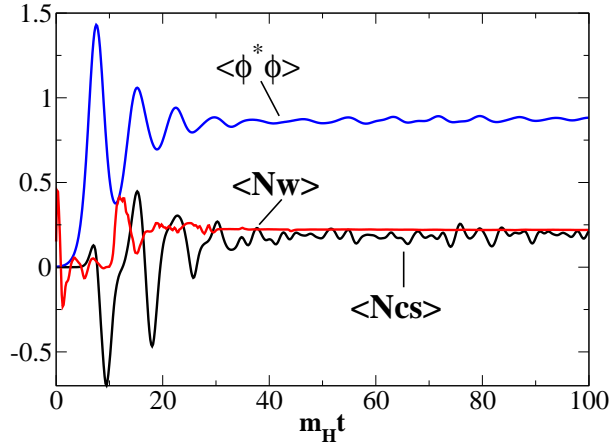}
}
\caption{Averages: Higgs field $2\langle\vrd\vr\rangle/v^2$, Chern-Simons number
$\langle\ncs\rangle$ and Higgs winding number $\langle \nw\rangle$, for $m_H=m_w$, $k=16$,
`just a half' initial conditions. }
\label{avtraj}
\end{figure}

We simulated with parameters such that $m_H/m_W = 1$ and $\sqrt{2}$, and various 
values of $k$ defined in 
(\ref{kappa}). The initial conditions were `just a half' (cf.\ Sect.\ \ref{quench})
and `thermal'. The latter are simply Higgs-field realizations drawn from a free
BE ensemble at low temperature $T=0.1\, m_H$, thermal noise 
that we used to get an impression of the sensitivity to the initial conditions.
Figure \ref{avtraj} shows an example of how at first
$\langle\vrd\vr\rangle$ rises exponentially and then settles into a damped
oscillation near the vev. 
The Chern-Simons number has a small bump near $t m_H = 7$, 
which we can understand analytically, after which
it appears to resonate with $\langle\vrd\vr\rangle$ and settle near the
winding number. The latter is at first erratic but then appears to stabilize
earlier than $\ncs$. 

\begin{figure}[h!]
\centerline{
\includegraphics*[width=8cm,clip]{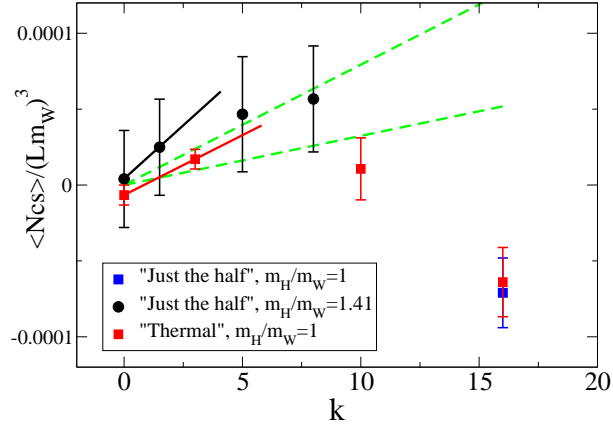}
}
\caption{Chern-Simons number density vs.\ $k$.}
\label{allkandm}
\end{figure}

Fig.\ \ref{allkandm} shows results for the Chern-Simons number-density at time
$t m_H = 100$. There is no evidence for a sensitive dependence on the initial
ensemble, and apparently the dependence on the magnitude of CP violation
becomes non-linear for $k\gtrsim 5-10$.
The dependence on the mass ratio 
$m_H/m_W$ appears limited. 
This is in contrast to what we found\cite{Smit:2002yg}
in the 1+1D abelian-Higgs toy model, for which there is a sensitive dependence on
$m_H/m_W$; it also showed nice linear behavior in the 
1+1D analog $\kp$ for small enough values. 
The dashed lines are linear fits through the origin, ignoring the data at $k=0$
and 16. The straight full lines go through the two data points with lowest $k$.
Including the data at $k=0$
is meaningful because these were generated with the same sequence of
pseudo-random numbers as for 
$k=3$ ($m_H = m_W$) and $k=1.5$ ($m_H = \sqrt{2}\, m_W$).
(Note that the $k=0$ data is shifted somewhat away from zero for $m_H = m_W$).
Using the dashed-line fit for the case $m_H = \sqrt{2}\, m_W\simeq 114$ GeV 
gives a baryon asymmetry
\[
\frac{n_{B}}{n_{\gamma}} =
(0.46\pm 0.21) \times 10^{-2}\kappa\, m_{W}^{2}.
\]
To fit the observed $n_B/n_\gm \simeq 6.5\times 10^{-10}$, requires $\kp$ to be
\[
\kappa\approx\frac{1.4 \times 10^{-7}}{m_{W}^{2}}
\approx \frac{2.2\times 10^{-5}}{1{\rm TeV}^{2}}.
\]
The 1 TeV scale is reasonable if we interpret 
$\mathcal{L}_{\rm CP}$ as being due to physics beyond the ESM. 
On the other hand, if we boldly assume that the results make sense for the
ESM by taking the scale $M = m_W$, then the required $\dlcp$ turns out
to be of the order of $J$ in (\ref{J}):
\[
\dl_{\rm CP} \approx 0.7\times 10^{-5}.
\]
We consider this as very encouraging for further developing 
the scenario of cold electroweak baryogenesis.

\section*{Acknowledgments}
I would like to thank 
Anders Tranberg, Jon-Ivar Skullerud and Bartjan van Tent, for a very fruitful 
and pleasant collaboration.
This work was supported by FOM/NWO.



\end{document}